\documentclass[11pt,a4paper]{article}
\usepackage{jheppub}
\usepackage{amsmath}
\usepackage[most]{tcolorbox}
\usepackage{dsfont}
\usepackage{ulem}
\usepackage{natbib}
\usepackage{xcolor}
\usepackage[hang,flushmargin]{footmisc}
\usepackage{tikz-cd}
\usepackage{enumitem}
\usepackage{amsfonts,amssymb, amscd,amsmath,latexsym,amsbsy,bm}
\usepackage{stmaryrd}
\usepackage{todonotes}
\usepackage{stackrel}

\usepackage{graphicx}

\usepackage[rightcaption]{sidecap}

\makeatletter\renewcommand{\@biblabel}[1]{#1.}\makeatother

\newtcolorbox{empheqboxed}{colback=gray!20, 
	colframe=white,
	width=\textwidth,
	sharpish corners,
	top=0mm, 
	bottom=0pt
}

\title{Lens partition function, pentagon identity\\ and star-triangle relation}

\author{Deniz N. Bozkurt$^a$, Ilmar Gahramanov$^{b,c,d}$, and Mustafa Mullahasanoglu$^{b}$}
\affiliation{$^a$ Department of Physics, Koc University, 34450 Sariyer, Istanbul, Turkey\\[-0.5cm]
	
	$^b$ {Department of Physics, Bogazici University, 34342 Bebek, Istanbul, Turkey}\\[-0.5cm]
	
	$^{c}$ Institute of Radiation Problems, Azerbaijan National Academy of Sciences, \\ B.Vahabzade St. 9, AZ1143, Baku, Azerbaijan\\[-0.5cm]
	
	$^{d}$ Department of Mathematics, Khazar University,  Mehseti St. 41, AZ1096, Baku, Azerbaijan \\[-0.5cm]
}

\emailAdd{dbozkurt16@ku.edu.tr}
\emailAdd{ilmar.gahramanov@boun.edu.tr}
\emailAdd{mustafa.mullahasanoglu@boun.edu.tr}

\abstract{We study the three-dimensional lens partition function for $\mathcal N=2$ supersymmetric gauge dual theories on $S_b^3/\mathbb{Z}_r$ by using the gauge/YBE correspondence. This correspondence relates supersymmetric gauge theories to exactly solvable models of statistical mechanics. The equality of partition functions for the three-dimensional supersymmetric dual theories can be written as an integral identity for hyperbolic hypergeometric functions.  We obtain such an integral identity which can be written as the star-triangle relation for Ising type integrable models and as the integral pentagon identity. The latter represents the basic 2-3 Pachner move for triangulated 3-manifolds. A special case of our integral identity can be used for proving orthogonality and completeness relation of the Clebsch-Gordan coefficients for the self--dual continuous series of $U_q(osp(1|2))$.
}


\begin{document}
	\maketitle
	\flushbottom

	\section{Introduction}

	The recent progress in gauge/YBE correspondence has lead to remarkable connections between supersymmetric gauge theories, integrable models of statistical mechanics, and special functions. The main idea of the correspondence is that the supersymmetric duality for gauge theories leads to the integrability for spin lattice models, see \cite{Gahramanov:2017ysd,Yamazaki:2018xbx} for a review and references therein. This interplay between supersymmetric theories and integrable models enables us to generate new solutions to the star-triangle relation which is a special form of the Yang-Baxter equation\footnote{There are IRF and vertex models studied in the context of gauge/YBE correspondence, in the paper we will only discuss Ising-type models.}, see e.g. \cite{Spiridonov:2010em,Kels:2015bda,Gahramanov:2015cva,Kels:2017toi,Jafarzade:2017fsc,de-la-Cruz-Moreno:2020xop,Gahramanov:2016ilb}. The star-triangle relation is a sufficient condition for integrability of Ising-type lattice models \cite{Baxter:1982zz,Baxter:1997tn}. It seems that the gauge/YBE correspondence gives a general method to construct solutions to the star-triangle relation.
	
	In this work we use the gauge/YBE correspondence to obtain the star-triangle relation and the pentagon identity in terms of hyperbolic hypergeometric functions. From the gauge theory side we consider the partition functions of $\mathcal N=2$ supersymmetric dual gauge theories on $S_b^3/\mathbb{Z}_r$. Such lens partition functions were studied from different aspects in several papers, see, e.g. \cite{Gang:2019juz,Benini:2011nc,Imamura:2012rq,Imamura:2013qxa,Nieri:2015yia,Gahramanov:2016ilb,Alday:2012au,Yamazaki:2013fva,Eren:2019ibl,Honda:2016vmv,Nedelin:2016gwu}.
	
	As the main result of the paper one may regard the following hyperbolic hypergeometric identity, 
	\begin{align}
	\frac{1}{r\sqrt{-\omega_1\omega_2}}\sum_{y=0}^{[ r/2 ]}\epsilon (y) e^{\frac{\pi iC}{2}}\int _{-\infty}^{\infty} dz & \prod_{i=1}^3\gamma^{(2)}(-i(a_i-z)-i\omega_1(u_i- y);-i\omega_1r,-i\omega_1-i\omega_2) \nonumber \\
	& \times \gamma^{(2)}(-i(a_i-z)-i\omega_2(r-u_i+ y);-i\omega_2r,-i\omega_1-i\omega_2) \nonumber \\
	& \times\gamma^{(2)}(-i(b_i+z)-i\omega_1(v_i+ y);-i\omega_1r,-i\omega_1-i\omega_2) \nonumber
	\\
	& \times \gamma^{(2)}(-i(b_i+z)-i\omega_2(r-(v_i+y));-i\omega_2r,-i\omega_1-i\omega_2)\nonumber\\
	= & \prod_{i,j=1}^3\gamma^{(2)}(-i(a_i+ b_j)-i\omega_1(u_i+ v_j);-i\omega_1r,-i\omega_1-i\omega_2)\nonumber\\ & \gamma^{(2)}(-i(a_i+ b_j)-i\omega_2(r-u_i- v_j);-i\omega_2r,-i\omega_1-i\omega_2))\;,
	\label{mainintegral}
	\end{align}
	with the balancing conditions $\sum_{i=1}^3 a_i+b_i=\omega_1+\omega_2$ and $\sum_{i=1}^3 u_i+v_i=0$. For the exponential term, we have $C=-2y+(u_1+u_2+u_3-v_1-v_2-v_3)$, the $\epsilon(y)$ function is defined as $\epsilon(0)=\epsilon(\lfloor\frac{r}{2}\rfloor)=1$ and $\epsilon(y)=2$ otherwise.  The hyperbolic gamma functions is defined as 
	\begin{align}
	\gamma^{(2)}(z;\omega_{1},\omega_{2})=\exp{\left(-\int_{0}^{\infty}\frac{dx}{x}\left[\frac{\sinh{x(2z-\omega_{1}-\omega_{2})}}{2\sinh{(x\omega_{1})}\sinh{(x\omega_{2})}}-\frac{2z-\omega_{1}-\omega_{2}}{2x\omega_{1}\omega_{2}}\right]\right)} \;.
	\end{align}
	We obtain the identity (\ref{mainintegral}) from the equality of partition functions of supersymmetric dual theories on $S_b^3/\mathbb{Z}_r$.
	The intriguing physical interpretation of this integral identity is that it can be written as the star-triangle relation for a certain two-dimensional Ising-type statistical model, as well as the pentagon identity for a certain triangulated $3$-manifold. The integrable model based on the identity (\ref{mainintegral}) is a generalization of the Faddeev-Volkov model \cite{Bazhanov:2007mh,Bazhanov:2007vg} and a special case of the model can be found in \cite{Gahramanov:2016ilb}. Here we only construct the edge-interacting lattice spin model, however the IRF version of the model may also give an interesting integral identity. 
	
	The Euler's gamma function limit of the integral identity (\ref{mainintegral}) gives the known solution to the star-triangle relation \cite{Bazhanov:2007vg}, also can be written as the pentagon identity presented in \cite{Jafarzade:2018yei}. From supersymmetric gauge theory side, by taking such a limit ($r \rightarrow \infty$) one obtains the partition function of two-dimensional $\mathcal N=(2,2)$ supersymmetric gauge theories on two-sphere $S^2$. 
	
	A special case when $r=2$, the identity (\ref{mainintegral}) gives the star-triangle relation discussed in \cite{Hadasz:2013bwa} which was used for proving orthogonality and completeness relation of the Clebsch-Gordan coefficients for the self-dual continuous series of $U_q(osp(1|2))$. We expect an intimate relation between supersymmetric gauge theories on $S_b^3/\mathbb Z_r$, quantum groups $U_q(osp(1|2))$ and two-dimensional conformal field theory.

	Some results of the paper agree exactly with the work\footnote{The relation to the supersymmetric lens partition function was not discussed in \cite{Sarkissian:2018ppc}.} \cite{Sarkissian:2018ppc}, based on a different interpretation of the integral identity (\ref{mainintegral}). However our approach is based on the supersymmetric gauge theory computations.

	The main idea of the paper is to construct connections between several solutions of the star-triangle equation and the pentagon relations. The following diagram demonstrates the plan of the paper, pictorially.
	

	\begin{figure}[h]
		\includegraphics{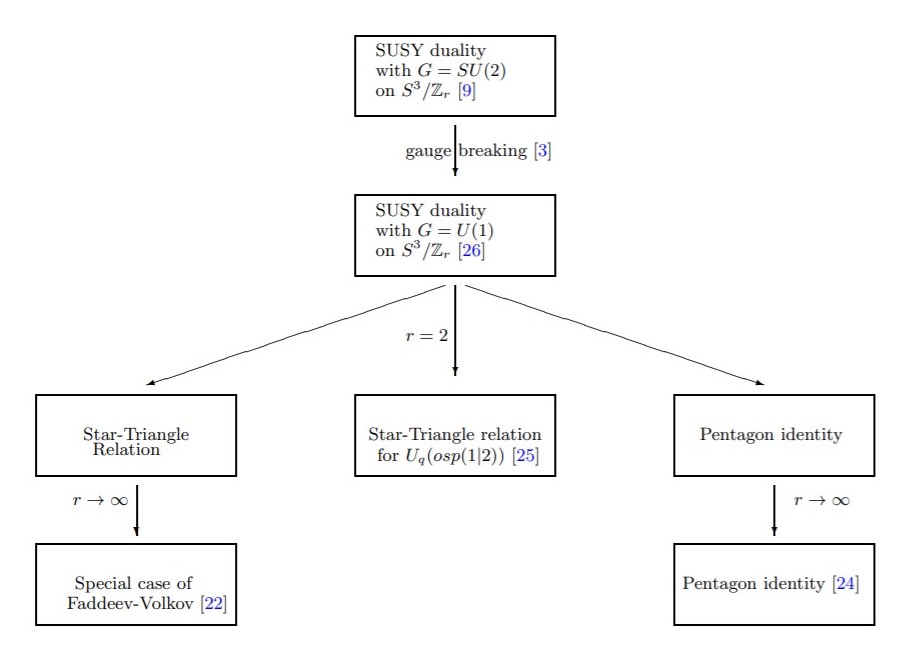}
		\caption{Structure of the paper.}	
	\end{figure}

	The rest of this paper is organized as follows. In section 2, we briefly recollect some basic definitions. In section 3, we present the star-triangle relations and pentagon identities resulting from the supersymmetric duality. In section 4, we discuss how to relate our star-triangle relation to the one obtained in \cite{Hadasz:2013bwa}. In section 5, we present our conclusions and discuss some open questions. We include three appendices for some technical details.

	\section{Lens partition function, 3d duality and gauge symmetry breaking}

	\subsection{Supersymmetric partition function on \texorpdfstring{$S^3_b/\mathbb{Z}_r$}{S3/Zr} }

	We start by defining the general form of the three dimensional $\mathcal N=2$ partition function on the squashed lens space $S^3_b/\mathbb{Z}_r$.  The lens partition function can be computed by a straightforward dimensional reduction of the four-dimensional lens superconformal index \cite{Benini:2011nc,Yamazaki:2013fva,Eren:2019ibl} or via the supersymmetric localization technique \cite{Imamura:2012rq,Imamura:2013qxa}. Here we briefly outline some basic ingredients\footnote{We mostly follow the notations of \cite{Gahramanov:2016ilb,Eren:2019ibl}.} and refer the reader to \cite{Imamura:2012rq,Imamura:2013qxa,Nieri:2015yia,Gahramanov:2016ilb} for more details. 
	
	Recall that the lens space $S_b^3/\mathbb Z_r$ can be obtained from the squashed three sphere 
	\begin{equation}
	S_b^3= \{(x,y)\in\mathbb{C}^2, b^2|x|^2+b^{-2}|y|^2=1\} \:,
	\end{equation}
	by making the identification $(x,y)\sim(e^{\frac{2\pi i}{r}}x,e^{\frac{2\pi i}{r}}y)$. The partition function on this manifold can be reduced to the following matrix model\footnote{Actually, this expression is the Coulomb branch localization result, one can get the partition function in different forms depending on the chosen localization locus \cite{Alday:2013lba,Willett:2016adv,Benini:2013yva}.}
	\begin{equation}
	Z=\sum_y\int \frac{1}{|W|}\prod_j^{rank G} \frac{dz_j}{2\pi i r}Z_{\text{classical}}[z,y]Z_{\text{one-loop}}[z,y] \;.
	\end{equation}
	Here the sum is over the holonomies $y=\frac{r}{2\pi}\int_C A_\mu dx^\mu$, where $C$ is the non-trivial cycle on $S^3_b/\mathbb{Z}_r$ and $A_\mu$ is the gauge field. The integral is over the Cartan subalgebra of the gauge group and $z_j$ variables are corresponding to Weyl weights. The order of the group G is represented by the prefactor $|W|$ such that the gauge group is broken by the holonomy into a product of $r$ subgroups. 
	
	The one-loop contribution of chiral multiplets is given in terms of hyperbolic hypergeometric function,
	\begin{equation}
	Z_{\rm chiral}=
	\prod_{j}\prod_{\rho_j}\prod_{\phi_j}
	\hat{s}_{b,-\rho_j(y)-\phi_j(n)}\left(i\frac{Q}{2}(1-\Delta_j)-\rho_j(z)-\phi_j(\Phi)\right)~.
	\end{equation}
	Here $j$ labels chiral multiplets;  $\rho_j,\phi_j$, are the weights of the representation of the   gauge  and  flavor groups, respectively and $\Delta_j$ is the Weyl weight of $j$'th chiral multiplet. We also define, $Q=b+\frac{1}{b}$ with the squashing parameter $b^2=\omega_2/\omega_1$. The function  $\hat{s}_{b,y}(z)$ is a version of the improved double sine function\footnote{Let us mention that  our $\hat{s}_b$ is different than one used in \cite{van2007hyperbolic,Amariti:2015vwa}.} \cite{Gahramanov:2016ilb}, which can be written as a product of hyperbolic gamma functions\footnote{Expressions in terms of the improved double sine function constitutes a special class of hyperbolic hypergeometric functions and they are in the interest of mathematical physics \cite{Gahramanov:2016ilb,Eren:2019ibl,Spiridonov:2016uae,Sarkissian:2018ppc,Cordova:2016jlu}.}
	\begin{align}
	\hat{s}_{b,-y}(x)=\sigma_h(y)\gamma^{(2)}(iz+y\omega_1+\eta;\omega_1r,2\eta)\gamma^{(2)}(iz+\omega_2(r-y)+\eta;\omega_2r,2\eta) \;,
	\label{s_b}
	\end{align}
	with $\sigma_h(y)=e^{\frac{i\pi}{2r}(y(r-y)-(r-1)y^2)}$ and $\eta=(\omega_1+\omega_2)/2$. For practical reasons, and in keeping with supersymmetric gauge theories notations, we will mainly use the hyperbolic gamma function $\gamma^{(2)}(z,\omega_1, \omega_2)$ instead of $\hat{s}_{b,y}(z)$. The one-loop contribution of the vector multiplet combined with the Vandermonde determinant can be written as
	\begin{align} \nonumber
	Z_{\rm vector} = \prod_{\alpha \in R_{+}} \frac{1}{\hat{s}_{b,\alpha(y)}\left(i\frac{Q}{2}+\alpha(z)\right)} \; ,
	\end{align}
	where the product is over the positive roots $\alpha$ of the gauge group $G$. Once we know the group-theoretical data of three-dimensional supersymmetric theory on $S_b^3/\mathbb Z_r$, we can write down the partition function in terms of hyperbolic hypergeometric integral. Note that in our examples the classical term $Z_{\text{classical}}$ which includes the contributions coming from classical action of the Chern-Simons term and Fayet-Iliopoulos term will be absent. We should mention that the expressions for multiplets are the same as the one used in \cite{Gahramanov:2016ilb,Eren:2019ibl} and differs by some factor (the partition function is the same) from that in \cite{Imamura:2012rq} and \cite{Nieri:2015yia}. The relation between two expressions can be found in \cite{Gahramanov:2016ilb} and in Appendix A.

	\subsection{Three-dimensional \texorpdfstring{$\mathcal N=2$}{N=2} IR duality}
	
	We will perform the gauge symmetry breaking of the following three-dimensional $\mathcal{N}=2$ supersymmetric dual theories\footnote{There is a four-dimensional version of this duality, see, e.g. \cite{Spiridonov:2008zr,Gahramanov:2013xsa}. The four-dimensional $\mathcal N=1$ theory also has $N_f=3$ flavors and usually dimensional reduction shifts the number of flavors by one. By adding a proper superpotential \cite {Aharony:2013dha} one can obtain a duality with the same number of flavors as in four dimensions.}:
	\begin{itemize}
		\item Theory \textbf{A}, has gauge group\footnote{Actually this duality is a special case of the family of dualities for the gauge group $SP(2 N_c)$. For $N_c=1$ this duality coincides with the $SU(2)$ \cite{Karch:1997ux,Dolan:2008qi}.} $SU(2)$ and flavor group $SU(6)$. The chiral multiplets transform under the fundamental representation of the gauge group and the flavor group; the vector multiplet transforms under the adjoint representation of the gauge group.

		\item  Theory \textbf{B}, is the dual to the theory  \textbf{A} without gauge symmetry. There are fifteen chiral multiplets in the totally antisymmetric tensor representation of the flavor group.  In our case, Theory \textbf{B} is the low energy description of Theory \textbf{A} which can be characterized purely by composite gauge singlets. 
		
	\end{itemize}
	
	Because of the supersymmetric duality one obtains the following equality of the partition functions\footnote{We will not go into details of the evaluation of partition functions for these dual theories, see \cite{Gahramanov:2016ilb} and references therein.}  \cite{Gahramanov:2016ilb}
	\begin{align} \label{SU2identity}
	\frac{1}{2r\sqrt{-\omega_1\omega_2}} \sum_{y=0}^{[ r/2 ]}\epsilon (y) \int _{-\infty}^{\infty} dz & \frac{\prod_{i=1}^6\gamma^{(2)}(-i(a_i\pm z)-i\omega_1(u_i\pm y);-i\omega_1r,-i\omega_1-i\omega_2)}{\gamma^{(2)}(\mp 2iz\mp 2i\omega_1y;-i\omega_1r,-i\omega_1-i\omega_2)} \nonumber \\
	\times & \frac{\gamma^{(2)}(-i(a_i\pm z)-i\omega_2(r-(u_i\pm y));-i\omega_2r,-i\omega_1-i\omega_2)}{\gamma^{(2)}(\mp2iz-i\omega_2(r\mp2y);-i\omega_2r,-i\omega_1-i\omega_2)} \nonumber \\ =\prod_{1\leq i<j\leq 6} & \gamma^{(2)}(-i(a_i + a_j)-i\omega_1(u_i + u_j);-i\omega_1r,-i\omega_1-i\omega_2) \nonumber \\ & 
	\gamma^{(2)}(-i(a_i + a_j)-i\omega_2(r-(u_i + u_j));-i\omega_2r,-i\omega_1-i\omega_2)\; ,
	\end{align}
	with the balancing conditions $\sum^6_{i=1} a_i=\omega_1+\omega_2$ and\space$\sum_{i=1}^6 u_i=0$. We should mention that there is a contribution of the $R$-symmetry appearing in the partition function but we absorbed it in the flavor fugacity. Since all physical degrees of freedom of Theory \textbf{B} are gauge invariant there is no summation and integration on the right hand side of the identity. The case $r=1$ of the integral identity (\ref{SU2identity}) is a very well-known integral identity, see, e.g. \cite{van2007hyperbolic}, in this case it corresponds to the equality of the squashed three-sphere partition functions \cite{Hama:2011ea}.  
	
	The hyperbolic hypergeometric beta sum-integral (\ref{SU2identity}) is an important identity in the theory of hyperbolic hypergeometric functions. Its role in integrable models of statistical mechanics was discovered in \cite{Gahramanov:2016ilb}.

	\subsection{Gauge Symmetry Breaking}
	
	Now we are in a position to obtain new dual theories by breaking the gauge symmetry. The idea is to break the gauge symmetry from $SU(2)$ to $U(1)$ in dual theories presented above. We give a VEV to two flavor quarks, breaking the gauge group to $U(1)$ and reducing the flavor group to $SU(3) \times SU(3)$. As a result we obtain the following dual theories:
	
	\begin{itemize}
		
		\item \textbf{Theory A:} $3d$ $\mathcal N=2$ theory with $U(1)$ gauge symmetry and $SU(3)_L \times SU(3)_R$ flavor group, chiral multiplets are belonging to the $SU(3)_L$ transforming in the fundamental representation of the gauge group and chiral multiplets are belonging to the $SU(3)_R$, transforming in the anti-fundamental representation. 
		
		\item \textbf{Theory B:} The dual theory has the same global symmetries without gauge degrees of freedom, nine ``mesons'', transforming in the fundamental representation of the flavor group $SU(3)_L \times SU(3)_R$.
		
	\end{itemize}
	
	We make the breaking of gauge symmetry on the level of partition functions. Following the work \cite{Spiridonov:2010em} (see also \cite{Sarkissian:2018ppc}) let us replace the flavor fugacities\footnote{In three dimensions it is a complexified real mass parameter.} $a_i$ to $a_i+\mu$ for $i=\{1,2,3\}$ and $a_i-\mu$ for $i=\{4,5,6\}$. We use the fact that the identity (\ref{SU2identity}) is symmetric with respect to $z\rightarrow -z$ transformation. By changing the variable $z$ to $z+\mu$ we get the following expression
	\begin{align} \nonumber
	\frac{1}{r\sqrt{-\omega_1\omega_2}}\sum_{y=0}^{[ r/2 ]}\epsilon (y) \int _{-\mu}^{\infty} dz & \frac{\prod_{i=1}^3\gamma^{(2)}(-i(a_i+\mu\pm (z+\mu))-i\omega_1(u_i\pm y);-i\omega_1r,-i\omega_1-i\omega_2)}{\gamma^{(2)}(\mp 2i(z+\mu)\mp 2i\omega_1y;-i\omega_1r,-i\omega_1-i\omega_2)} \nonumber \\
	\times & \frac{\gamma^{(2)}(-i(a_i+\mu\pm (z+\mu))-i\omega_2(r-(u_i\pm y));-i\omega_2r,-i\omega_1-i\omega_2)}{\gamma^{(2)}(\mp2i(z+\mu)-i\omega_2(r\mp2y);-i\omega_2r,-i\omega_1-i\omega_2)}  \nonumber \\
	\prod_{i=4}^6 & \gamma^{(2)}(-i(a_i-\mu\pm (z+\mu))-i\omega_1(u_i\pm y);-i\omega_1r,-i\omega_1-i\omega_2) \nonumber
	\\
	\times & \gamma^{(2)}(-i(a_i-\mu\pm (z+\mu))-i\omega_2(r-(u_i\pm y));-i\omega_2r,-i\omega_1-i\omega_2) \nonumber \\
	= \prod_{i=1}^3\prod_{j=4}^6 & \gamma^{(2)}(-i(a_i+ a_j)-i\omega_1(u_i+ u_j);-i\omega_1r,-i\omega_1-i\omega_2)\nonumber\\ \times & \gamma^{(2)}(-i(a_i+ a_j)-i\omega_2(r-(u_i+ u_j));-i\omega_2r,-i\omega_1+\omega_2) \nonumber \\   \prod_{1\leq i<j\leq 3} & \gamma^{(2)}(-i(a_{i+3}+a_{j+3}-2\mu)-i\omega_2(r-(u_{i+3}+u_{j+3}));-i\omega_2r,-i\omega_1-i\omega_2)
	\nonumber \\\times & \gamma^{(2)}(-i(a_i+a_j+2\mu)-i\omega_2(r-(u_i+u_j));-i\omega_2r,-i\omega_1-i\omega_2) \nonumber \\
	\times & \gamma^{(2)}(-i(a_{i+3}+a_{j+3}-2\mu)-i\omega_1(u_{i+3}+u_{j+3});-i\omega_1r,-i\omega_1-i\omega_2) \nonumber \\\times &\gamma^{(2)}(-i(a_i+a_j+2\mu)-i\omega_1(u_i+u_j);-i\omega_1r,-i\omega_1-i\omega_2) \:.
	\end{align}

	After taking the limit $\mu\to\infty$ and renaming the flavor group coefficients as $a_i=b_i$ and $u_i=v_i$ for only $i\in\{4,5,6\}$, the reduced form of the hyperbolic hypergeometric integral identity turns out to be
	
	\begin{align}
	\frac{1}{r\sqrt{-\omega_1\omega_2}}\sum_{y=0}^{[ r/2 ]}\epsilon (y) e^{\frac{\pi iC}{2}}\int _{-\infty}^{\infty} dz\prod_{i=1}^3 & \gamma^{(2)}(-i(a_i-z)-i\omega_1(u_i- y);-i\omega_1r,-i\omega_1-i\omega_2) \nonumber \\
	\times & \gamma^{(2)}(-i(a_i-z)-i\omega_2(r-(u_i- y));-i\omega_2r,-i\omega_1-i\omega_2) \nonumber \\
	\times & \gamma^{(2)}(-i(b_i+z)-i\omega_1(v_i+ y);-i\omega_1r,-i\omega_1-i\omega_2) \nonumber
	\\
	\times & \gamma^{(2)}(-i(b_i+z)-i\omega_2(r-(v_i+y));-i\omega_2r,-i\omega_1-i\omega_2)\nonumber\\
	= \prod_{i,j=1}^3& \gamma^{(2)}(-i(a_i+ b_j)-i\omega_1(u_i+ v_j);-i\omega_1r,-i\omega_1-i\omega_2)\nonumber\\ \times& \gamma^{(2)}(-i(a_i+ b_j)-i\omega_2(r-(u_i+ v_j));-i\omega_2r,-i\omega_1-i\omega_2))\;, \label{mainintegraltext}
	\end{align}
	with the balancing conditions $\sum_{i=1}^3 a_i+b_i=\omega_1+\omega_2$ and $\sum_{i=1}^3 u_i+v_i=0$. Here $C=-2y+(u_1+u_2+u_3-v_1-v_2-v_3)$. 
	On the left hand-side of the integral identity, we see the partition function of the Theory \textbf{A} and on the right hand-side, the Theory \textbf{B}. A similar identity was discussed for the $S_b^3$ sphere partition functions in \cite{Spiridonov:2010em,Kashaev:2012cz} and for the superconformal indices in \cite{Gahramanov:2013rda,Gahramanov:2014ona,Gahramanov:2016wxi}. The integral identity (\ref{mainintegraltext}) is essentially the same integral identity as the one obtained in \cite{Sarkissian:2018ppc} with a slightly different sign coefficient.

	\section{Star-triangle relation and pentagon identity}
	
	In this section we investigate the relation between supersymmetric dualities, integrability and triangulated 3-manifolds. We will show that the integral identity (\ref{mainintegraltext}) can be written as the star-triangle relation and as the integral pentagon identity.

	\subsection{Pentagon Identity}
	
	The integral identity (\ref{mainintegraltext}) can be written as a pentagon relation. The pentagon identity usually represents the basic 2-3 Pachner move \cite{pachner1991pl} for a certain triangulated 3-manifold. There are several examples of integral pentagon relations computed via three-dimensional supersymmetric dualities, see, e.g. \cite{Dimofte:2011ju,Kashaev:2012cz,Gahramanov:2013rda, Gahramanov:2014ona, Gahramanov:2016wxi, Benvenuti:2016wet, Bozkurt:2018xno, Jafarzade:2018yei}. Here we present a new pentagon identity in terms of hyperbolic gamma functions.
	
	It is convenient to define the following function
	\begin{align}
	\mathcal{B}(z_1,u_1;z_2,u_2)=\frac{\gamma^{(2)}(-iz_1-i\omega_1u_1;-i\omega_1r,-i\omega_1-i\omega_2)\gamma^{(2)}(-iz_1-i\omega_2(r-u_1);-i\omega_2r,-i\omega_1-i\omega_2)}{\gamma^{(2)}(-i(z_1+z_2)-i\omega_1(u_1+u_2);-i\omega_1r,-i\omega_1-i\omega_2)} \nonumber \\\frac{\gamma^{(2)}(-iz_2-i\omega_1u_2;-i\omega_1r,-i\omega_1-i\omega_2)\gamma^{(2)}(-iz_2-i\omega_2(r-u_2);-i\omega_2r,-i\omega_1-i\omega_2)}{\gamma^{(2)}(-i(z_1+z_2)-i\omega_2(r-u_1-u_2);-i\omega_2r,-i\omega_1-i\omega_2)} \; ,
	\label{pentagon_b}
	\end{align}
	which solves the following integral pentagon identity,\footnote{One can think that our pentagon relation coincides with the one obtained in \cite{Alexandrov:2015xir}. However they are different, actually the identity (3.10) (or (4.12)) in \cite{Alexandrov:2015xir} can be obtained from the three-dimensional $\mathcal N=2$ mirror symmetry on $S_b^3/\mathbb Z_k$ for special values of flavor fugacities, see \cite{Imamura:2012rq}.}
	\begin{align}
	\frac{1}{r\sqrt{-\omega_1\omega_2}}\sum_{y=-\lfloor r/2 \rfloor}^{\lfloor r/2 \rfloor} e^{\frac{\pi iC}{2}} \int _{-\infty}^{\infty} dz \prod_{i=1}^3\mathcal{B}(a_i-z,u_i-y;b_i+z,v_i+y)\nonumber \\=\mathcal{B}(a_1+b_2,u_1+v_2;a_2+b_3,u_2+v_3)\mathcal{B}(a_1+b_3,u_1+v_3;a_2+b_1,u_2+v_1) 
	\label{pentagon}
	\end{align}
	with the same balancing conditions given in (\ref{mainintegraltext}).
	
	\subsection{Limit of the Pentagon Identity}
	
	There are several pentagon identities in terms of Euler's gamma function \cite{Kashaev:2012cz,kashaev2014euler,Jafarzade:2018yei}. Here we present the pentagon relation found in \cite{Jafarzade:2018yei} in terms of gamma function which we obtain in a different manner\footnote{The derivation of the pentagon identity in \cite{Jafarzade:2018yei} is based on the reduction procedure of the three-dimensional $\mathcal N=2$ superconformal index (the partition function on $S^2\times S^1$) to the $\mathcal N=(2,2)$ supersymmetric sphere partition function by shrinking the radius of $S^1$ \cite{Benini:2012ui}.}.
	
	In order to explore the limit of the pentagon identity we use the following asymptotic relation
	\begin{equation}
	\lim_{\omega_2\to\infty} \Big(\frac{\omega_2}{2\pi\omega_1}\Big)^{\frac{z}{\omega_2}-\frac{1}{2}}\gamma^{(2)}(z;\omega_1,\omega_2)=\frac{\Gamma(z/\omega_1)}{\sqrt{2\pi}} \;.
	\label{gamma_limit}
	\end{equation}
	We identify $\omega_1$ with $\omega_2$ and redefine all coefficients dividing by $\frac{\omega_1+\omega_2}{\omega_1}$, then by altering $\frac{a_i-z}{\omega_1}$ to $a_i-z$, we obtain the limit of the pentagon identity as follows
	\begin{align}
	\sum_{y=-\infty}^{\infty}  \int _{-\infty}^{\infty} \frac{dz}{4\pi i} \prod_{i=1}^3\frac{\Gamma\left(a_i-z+\frac{u_i- y}{2}\right)
		\Gamma\left(b_i+z+\frac{v_i+ y}{2}\right) 
		\Gamma\left(1-a_i- b_i+\frac{u_i+ v_i}{2}\right) }{\Gamma\left(1-b_i-z+\frac{v_i+y}{2}\right)\Gamma\left(1-a_i+z+\frac{u_i- y}{2}\right)\Gamma\left(a_i+ b_i+\frac{u_i+ v_i}{2}\right)}\nonumber \\=\prod_{i,j=1; i\neq j}^3\frac{\Gamma\left(a_i+ b_j+\frac{u_i+ v_j}{2}\right)}{\Gamma\left(1-a_i- b_j+\frac{u_i+ v_j}{2}\right)} \;.
	\end{align}
	If we introduce the following function
	\begin{align}
	\mathcal{B}(z_1,u_1;z_2,u_2)=\frac{\Gamma\left(z_1+\frac{u_1}{2}\right)\Gamma\left(z_2+\frac{u_2}{2}\right)\Gamma\left(1-z_1-z_2-\frac{u_1+u_2}{2}\right)}{\Gamma\left(z_1+z_2+\frac{u_1+u_2}{2}\right)\Gamma\left(1-z_1-\frac{u_1}{2}\right)\Gamma\left(1-z_2-\frac{u_2}{2}\right)} \; ,
	\end{align}
	one obtains the integral pentagon identity
	\begin{align}
	\sum^{\infty}_{y=-\infty}\int_{-\infty}^\infty \frac{dz}{4\pi i} \prod_{i=1}^3 & \mathcal{B}(a_i-z,u_i- y;b_i+z,v_i+ y)\nonumber \\
	= & \mathcal{B}(a_1+b_3,u_1+v_3;a_2+b_1,u_2+v_1)
	\mathcal{B}(a_1+b_2,u_1+v_2;a_2+b_3,u_1+v_3) \;. \label{limit_pentagon}
	\end{align}
	This is exactly the result obtained in \cite{Jafarzade:2018yei} via dimensional reduction of the three-dimensional $\mathcal N=2$ supersymmetric dual theories on $S^2 \times S^1$.

	\subsection{Star-triangle relation}
	
	The star-triangle relation is a crucial equation in the study of two-dimensional integrable lattice spin models. Here we obtain solution to the star-triangle relation mentioned in \cite{Sarkissian:2018ppc}.
	We fix the parameters as
	\begin{align}
	a_i & =-\alpha_i+x_{i}\;, ~~~~~~~~~~~~~\; b_i=-\alpha_i-x_{i} \;,
	\end{align}
	and we insert the condition $u_i=-v_i$. By defining the  Boltzmann weight as
	\begin{align}
	W_{\alpha}(x_i,x_j,u_i,u_j)=&e^{-\pi i(u_i+u_j)}\gamma^{(2)}(-i(-\alpha+x_i-x_j)-i\omega_1(u_i-u_j);-i\omega_{1}r,-i\omega_1-i\omega_2)\nonumber \\&\gamma^{(2)}(-i(-\alpha+x_i-x_j)-i\omega_2(r-(u_i-u_j));-i\omega_2r,-i\omega_1-i\omega_2)\nonumber \\&\gamma^{(2)}(-i(-\alpha-x_i+x_j)-i\omega_1(u_j-u_i);-i\omega_{1}r,-i\omega_1-i\omega_2)\nonumber \\&\gamma^{(2)}(-i(-\alpha-x_i+x_j)-i\omega_2(r-(u_j-u_i));-i\omega_2r,-i\omega_1-i\omega_2) \; 
	\label{boltzman_weight}
	\end{align}
	and the spin-independent weight as
	\begin{align}
	R(\alpha_1,\alpha_2,\alpha_3)=\prod_{j=1}^3\gamma^{(2)}(2i\alpha_j;-i\omega_{1}r,-i(\omega_1+\omega_{2}))\gamma^{(2)}(2i\alpha_j;-i\omega_{2}r,-i(\omega_1+\omega_{2}))\; ,
	\end{align}
	one can rewrite the integral identity (\ref{mainintegraltext}) as the following star-triangle relation 
	\begin{align}\nonumber
	\frac{1}{r\sqrt{-\omega_{1}\omega_{2}}}\sum_{y=-\lfloor r/2\rfloor}^{\lfloor r/2\rfloor}e^{-2\pi iy} \int_{-\infty}^\infty dz W_{\alpha_1}(x_{1},z,u_1,y)W_{\alpha_2}(x_{2},z,u_2,y)W_{\alpha_3}(x_{3},z,u_3,y) \\ =R(\alpha_1,\alpha_2,\alpha_3) W_{\alpha_1+\alpha_2}(x_{1},x_{2},u_1,u_2)W_{\alpha_1+\alpha_3}(x_{1},x_{3},u_1,u_3)W_{\alpha_2+\alpha_3}(x_{2},x_{3},u_2,u_3) \;.
	\label{str}
	\end{align}
	Our model is an Ising type model where sites of the lattice are assigned to discrete $y$ and continuous spin $z$ variables. 
	
	\subsection{Limit of the Star-Triangle Relation}
	
	There are several solutions to the star-triangle relation in terms of Euler's gamma function. In our case such a solution can be achieved by taking the limit (\ref{gamma_limit}). After taking the limit, we obtain the following Boltzmann weight
	\begin{align} \label{GammaBoltzmann}
	W_{\theta}(x,z,u,y)=\frac{\Gamma(\frac{1-\frac{\theta}{\pi}+ix-iz+u- y}{2})
		\Gamma(\frac{1-\frac{\theta}{\pi}-ix+iz+v+ y}{2}) }{\Gamma(\frac{1+\frac{\theta}{\pi}+ix-iz-u+y}{2})\Gamma(\frac{1+\frac{\theta}{\pi}-ix+iz-u+ y}{2})} \;,
	\end{align}
	and spin independent weight
	\begin{align}
	R(\theta_1,\theta_2,\theta_3)=\prod_{i=1}^3 \frac{\Gamma(1-\frac{\theta_i}{\pi})}{\Gamma(\frac{\theta_i}{\pi})} \; .
	\end{align}
	The star-triangle relation can be written as
	\begin{align}\nonumber
	\sum_{y=-\infty}^{\infty} \int_{-\infty}^\infty \frac{dz}{8\pi} W_{\theta_1}(x_{1},z,u_1,y)W_{\theta_2}(x_{2},z,u_2,y)W_{\theta_3}(x_{3},z,u_3,y) \\ =R(\theta_1,\theta_2,\theta_3) W_{\pi-\theta_1}(x_{2},x_{3},u_2,u_3)W_{\pi-\theta_2}(x_{1},x_{3},u_1,u_3)W_{\pi-\theta_3}(x_{1},x_{2},u_1,u_2) \; ,
	\label{limit_str}
	\end{align}
	where $\theta_1+\theta_2+\theta_3=2\pi$.
	It can be easily checked that this solution is exactly the one obtained in \cite{Bazhanov:2007vg} from the Faddeev-Volkov model. In \cite{Bazhanov:2007vg} authors normalized the Boltmann weights  (\ref{GammaBoltzmann}) in such a way that the spin-independent function $R(\theta_1, \theta_2, \theta_3)$ is equal to one. Note that the solution (\ref{GammaBoltzmann}) is related to the special case of the Zamolodchikov's ``fishnet'' model \cite{Zamolodchikov:1980mb,Bazhanov:2007vg,Bazhanov:2016ajm}.

	\section{Relation to the \texorpdfstring{$U_q(\text{osp}(1|2))$}{Uq(osp(1|2))}}
	
	It is well known that the unitary representations of the modular double of $U_q(sl(2, R))$ is equivalent to the representations of the Liouville theory. For instance, $3j$-symbols for the tensor product of modular double representations of $U_q(sl(2, R))$  appear in the fusion product for the Liouville vertex operators. The modular double representation for the  $U_q(osp(1|2))$ plays\footnote{It is the $q$-deformed universal enveloping algebra of the Lie superalgebra $osp(1|2)$ with the deformation parameter $q=e^{i\pi b^2}$ \cite{Kulish:1988gr,Kulish:1989sv,Saleur:1989gj}.} the same role in the $\mathcal N = 1$ supersymmetric Liouville theory.
	
	Here we show how one can obtain the star-triangle relation for the $U_q(osp(1|2))$ \cite{Pawelkiewicz:2013wga,Hadasz:2013bwa} (see also \cite{Poghosyan:2016kvd}) from the integral identity (\ref{mainintegraltext}). The computations presented here and in Appendix C overlap with the computations of \cite{Sarkissian:2018ppc}. We use different notations and present all calculations in detail, see Appendix C. From the supersymmetric gauge theory point of view, the star-triangle relation represents the equality of partition functions of dual three-dimensional $\mathcal N=2$ gauge theories on $S_b^3/Z_2$ (it is topologically $\mathbb{RP}^3$). Similar computations for the Liouville field theory and  the supersymmetric gauge theories on squashed three-sphere $S_b^3$ was performed in \cite{Teschner:2012em}.

	A special case of the expression (\ref{mainintegraltext}) when r = 2, gives the star-triangle relation discussed in \cite{Hadasz:2013bwa} which can be used for proving orthogonality and completeness relation of the Clebsch-Gordan coefficients for the self-dual continuous series of $U_q(osp(1|2))$ and the $\mathcal N = 1$ supersymmetric Liouville theory. For the special case $r=2$ we obtain the following expression\footnote{For details, see Appendix C.}
	\begin{align} \label{r=2iden}
	\sum_{\nu=0,1}(-1)^{\frac{2\nu-3-\sum_i( \mu_i-\nu_i)}{2}} \int\frac{dx}{i}\prod_{i=1}^3S_{\nu+\nu_i}(x+a_i)S_{1+\nu+\mu_i}(b_i-x)=2\prod_{i,j=1}^3S_{1+\nu_i+\mu_j}(a_i+b_j) \;,
	\end{align}
	where we introduced the notations of the work \cite{Hadasz:2013bwa}
	\begin{equation}
	S_\nu(x)=\gamma^{(2)}(\frac{x+(1-\nu)b}{2},b,\frac{1}{b}) \gamma^{(2)}(\frac{x+1/b+\nu b}{2},b,\frac{1}{b}) \;.
	\end{equation}
	Note that we have a different sign coefficient\footnote{In \cite{Hadasz:2007wi}, the identity is given in the form
		\begin{align}
		\sum_{\nu=0,1}(-1)^{\nu(1+\sum_i(\nu_i+\mu_i))/2}\int\frac{dx}{i}\prod_{i=1}^3S_{\nu+\nu_i}(x+a_i)S_{1+\nu+\mu_i}(b_i-x)=2\prod_{i,j=1}^3S_{1+\nu_i+\mu_j}(a_i+b_j)\nonumber \;.
		\end{align}} in (\ref{r=2iden}) than in \cite{Hadasz:2013bwa,Sarkissian:2018ppc}. It seems that one can obtain the integral identity (5.14) from the work  \cite{Hadasz:2007wi} by tending one of the flavor fugacities to $b+\frac{1}{b}$.
	
	\section{Conclusion}
	
	We obtain the pentagon identity (\ref{pentagon}) (related to the Heisenberg double) and the star-triangle relation (\ref{str}) (related to the quantum algebra) from the same integral identity. Note that it is possible to construct the Boltzmann weight $W$ (\ref{boltzman_weight}) from the $B$-function (\ref{pentagon_b}), see e.g. \cite{kashaev1997heisenberg,Faddeev:1999fe,hikami2001hyperbolic}. 
	
	There are several directions that we wish to pursue in the future. We showed that, by performing a suitable identification, our star-triangle relation gives the same identity obtained in \cite{Hadasz:2013bwa}. This result is interesting not only because it builds a relation between two different subjects, but also it can be applied to arbitrary $r$. The problem we leave to the future is the construction of the corresponding quantum algebra for  the integral identity (\ref{limit_str}) with the general $r$.  
	
	In this work we presented rational and trigonometric solutions to  the Yang–Baxter equation in the form called star-triangle relation. It would be interesting to construct the operator form of the Yang-Baxter equation and the Hamiltonian of the one-dimensional chain corresponding to this solution.
	
	The pentagon identity and the star-triangle relation are a consequence of the Heisenberg double and the quantum algebra, respectively. We should mention that the appearance of the pentagon relation refers to the Pachner's move for triangulated 3-manifolds, though, we do not know how to construct this relation formally.

	\section*{Acknowledgements}
	
	It is a pleasure to acknowledge Ege Eren and Shahriyar Jafarzade for our collaboration at an early stage of this work. We would like to thank Michal Pawelkiewicz  for sharing his notes and his thesis. The work of Ilmar Gahramanov is partially supported by the Bogazici University Research Fund under grant number 20B03SUP3 and by the BAP Project (no. 2019-26) funded by Mimar Sinan Fine Arts University. Mustafa Mullahasanoglu is supported by the 2209-TUBITAK National/International Research Projects Fellowship Programme for Undergraduate Students under grant number 1919B011902237. 

	\appendix
	
	\section{Properties of special functions}
	
	
	Here we present several definitions and notations of special functions needed in this work.
	
	We briefly summarize the basic properties of hyperbolic gamma function and its different notations \cite{van2007hyperbolic, Andersen:2014aoa}. This function appears in several areas of mathematical and theoretical physics, here is an incomplete list of these topics.
	\begin{itemize}
		\item knot theory \cite{hikami2001hyperbolic,Hikami_2007,hikami2014braiding,Chan:2017qnw}
		\item supersymmetric gauge theory \cite{Teschner:2012em}
		\item integrable models of statistical mechanics \cite{Bazhanov:2007mh,Bazhanov:2007vg}
		\item special functions \cite{van2007hyperbolic}
	\end{itemize}

	\par With parameters $\tilde{q}=e^{2\pi i \omega_{1}/\omega_{2}}$ and $q=e^{-2\pi i \omega_{2}/\omega_{1}}$, the infinite product representation is
	\begin{align}
	\gamma^{(2)}(z;\omega_{1},\omega_{2})=e^{\frac{\pi i}{2}B_{2,2}(z;\omega_{1},\omega_{2})}\frac{(e^{-2\pi i\frac{z}{\omega_{2}}}\tilde{q};\tilde{q})}{(e^{-2\pi i\frac{z}{\omega_{1}}};q)} \; ,
	\end{align}
	where we have one of the Bernoulli polynomials,
	\begin{align}
	B_{2,2}(z;\omega_{1},\omega_{2})=\frac{z^2-z(\omega_{1}+\omega_{2})}{\omega_{1}\omega_{2}}+\frac{\omega_{1}^2+3\omega_{1}\omega_{2}+\omega_{2}^2}{6\omega_{1}\omega_{2}}.
	\end{align}
	Here, we realise that $B_{2,2}(z;\omega_{1},\omega_{2})$ is crucial for the asymptotic behavior of the hyperbolic gamma function. The hyperbolic gamma function has an integral representation\footnote{Actually, there are several integral representations, see, e.g. \cite{Faddeev:1995nb,woronowicz2000quantum}.} 
	\begin{align}
	\gamma^{(2)}(z;\omega_{1},\omega_{2})=\exp{\left(-\int_{0}^{\infty}\frac{dx}{x}\left[\frac{\sinh{x(2z-\omega_{1}-\omega_{2})}}{2\sinh{(x\omega_{1})}\sinh{(x\omega_{2})}}-\frac{2z-\omega_{1}-\omega_{2}}{2x\omega_{1}\omega_{2}}\right]\right)} \; ,
	\end{align}
	where $Re(\omega_{1}),Re(\omega_{2})>0$ and $Re(\omega_{1}+\omega_{2})>Re(z)>0$. \\
	We list here some properties of this function.
	\begin{align}
	\text{Symmetry:} & ~~~ & \gamma^{(2)}(z;\omega_{1},\omega_{2})=\gamma^{(2)}(z;\omega_{2},\omega_{1})  \\
	\text{Reflection:} & ~~~ & \gamma^{(2)}(z;\omega_{1},\omega_{2})\gamma^{(2)}(\omega_{1}+\omega_{2}-z;\omega_{1},\omega_{2})=1 \\
	\text{Scaling:} & ~~~ & \gamma^{(2)}(z;\omega_{1},\omega_{2})=\gamma^{(2)}(u z;u\omega_{1},u\omega_{2})\\
	\text{Conjugation:} & ~~~ & \gamma^{(2)}(z;\omega_{1},\omega_{2})^{*}=\gamma^{(2)}(z^{*};\omega_{2}^{*},\omega_{1}^{*})
	\end{align}
	Another very important property of the hyperbolic gamma function is the following difference equation
	\begin{equation}
	\gamma^{(2)}(z+\omega_1;\omega_1,\omega_2)=e^{-\frac{\pi i}{2}\left(B_{2,2}(z+\omega_1;\omega)-B_{2,2}(z;\omega)\right)}(1-e^{2\pi i\frac{z}{\omega_{2}}})\gamma^{(2)}(z;\omega_1,\omega_2) \; ,
	\end{equation}
	after simplifying, the difference equation takes the form,
	\begin{equation}
	\gamma^{(2)}(z+\omega_1;\omega_1,\omega_2)=2\sin\left(\frac{\pi z}{\omega_2}\right)\gamma^{(2)}(z;\omega_1,\omega_2) \; .
	\end{equation}
	Now we introduce the asymptotic behaviour of the function
	\begin{align}
	\lim_{z\to\infty}e^{\frac{\pi i}{2}B_{2,2}(z;\omega_{1},\omega_{2})}\gamma^{(2)}(z;\omega_{1},\omega_{2})=1\: \: \text{for} \:  \: \arg{\omega_{2}+\pi}>\arg{z}>\arg{\omega_{1}}   \\
	\lim_{z\to\infty}e^{-\frac{\pi i}{2}B_{2,2}(z;\omega_{1},\omega_{2})}\gamma^{(2)}(z;\omega_{1},\omega_{2})=1 \: \: \text{for} \: \: \arg{\omega_{2}}>\arg{z}>\arg{\omega_{1}-\pi} \; ,
	\end{align}
	where $\text{Im}(\frac{\omega_{1}}{\omega_{2}})>0$. We use these formulas for the breaking of gauge symmetry given in Appendix B.

	There is a generalization of the hyperbolic gamma function $\Gamma_h(z,y;\omega_1,\omega_2)$ which was introduced in \cite{Gahramanov:2016ilb}. This function can be defined in terms of $\gamma^{(2)}(z;\omega_1,\omega_2)$ as follows
	\begin{equation}
	\Gamma_h(z,y;\omega_1,\omega_2)=e^{\phi(y)}\gamma^{(2)}(-iz-i\omega_1y;-i\omega_1r,-i(\omega_1+\omega_2))\gamma^{(2)}(-iz-i\omega_2(r-y);-i\omega_2r,-i(\omega_1+\omega_2)) \; ,
	\label{gamma_h}
	\end{equation}
	where $\phi(y)=-\frac{\pi i}{6r}(2y^3-3y^2r+yr^2)$. It has the following properties.
	\begin{align}
	\text{Symmetry and Reflection:} & ~~~ & \Gamma_{h}(z,y;\omega_{1},\omega_{2})\Gamma_{h}(\omega_{1}+\omega_{2}-z,y;\omega_{2},\omega_{1})=e^{2\phi(y)} \\
	\text{Scaling:} & ~~~ & \frac{\Gamma_{h}(uz,uy;u\omega_{1},u\omega_{2})}{\Gamma_{h}(z,y;\omega_{2},\omega_{1})}=e^{\phi(uy)-\phi(y)}\\
	\text{Conjugation:} & ~~~ & \Gamma_{h}(z,y;\omega_{1},\omega_{2})^{*}=\Gamma_{h}(z^{*},y^{*};\omega_{2}^{*},\omega_{1}^{*}) 
	\end{align}
	
	\section{Gauge Symmetry Breaking}
	We start by reparemetrizing the integral identity coming from the duality argument, given in (\ref{mainintegraltext})
	\begin{align} \nonumber
	\frac{1}{2r\sqrt{-\omega_1\omega_2}} \sum_{y=0}^{[ r/2 ]}\epsilon (y) \int _{-\infty}^{\infty} dz\frac{\prod_{i=1}^6\gamma^{(2)}(-i(a_i\pm z)-i\omega_1(u_i\pm y);-i\omega_1r,-i(\omega_1+\omega_2))}{\gamma^{(2)}(\mp 2iz\mp i\omega_1(2y);-i\omega_1r,-i(\omega_1+\omega_2))} \nonumber \\
	\times \frac{\gamma^{(2)}(-i(a_i\pm z)-i\omega_2(r-(u_i\pm y));-i\omega_2r,-i(\omega_1+\omega_2))}{\gamma^{(2)}(\mp2iz-i\omega_2(r\mp(2y));-i\omega_2r,-i(\omega_1+\omega_2))} \nonumber \\ =\prod_{1\leq i<j\leq 6}\gamma^{(2)}(-i(a_i + a_j)-i\omega_1(u_i + u_j);-i\omega_1r,-i(\omega_1+\omega_2)) \nonumber \\ \gamma^{(2)}(-i(a_i + a_j)-i\omega_2(r-(u_i + u_j));-i\omega_2r,-i(\omega_1+\omega_2))\; ,
	\end{align}
	with the balancing condition $\sum^6_{i=1} a_i=\omega_1+\omega_2$ and\space$\sum_{i=1}^6 u_i=0$.
	
	As we add $\mu$ to first three coefficients and z variable, coming from the fundamental representation of the flavor group and gauge group; subtract $\mu$ from the last three coefficients, the left hand side of the equation turns out to be
	\begin{align} 
	\frac{1}{r\sqrt{-\omega_1\omega_2}}\sum_{y=0}^{[ r/2 ]}\epsilon (y) \int _{-\mu}^{\infty} dz\prod_{i=1}^3\gamma^{(2)}(-i(a_i-z)-i\omega_1(u_i- y);-i\omega_1r,-i(\omega_1+\omega_2)) \nonumber \\
	\times \gamma^{(2)}(-i(a_i-z))-i\omega_2(r-(u_i- y));-i\omega_2r,-i(\omega_1+\omega_2))  \nonumber \\
	\times\prod_{i=4}^6\gamma^{(2)}(-i(a_i+z)-i\omega_1(u_i+ y);-i\omega_1r,-i(\omega_1+\omega_2)) \nonumber
	\\
	\times \gamma^{(2)}(-i(a_i+z)-i\omega_2(r-(u_i+ y));-i\omega_2r,-i(\omega_1+\omega_2)) \nonumber \\
	\Bigg[\frac{\prod_{i=1}^3\gamma^{(2)}(-i(a_i+z+2\mu)-i\omega_1(u_i+y);-i\omega_1r,-i(\omega_1+\omega_2))}{\gamma^{(2)}(\mp 2i(z+\mu)\mp i\omega_1(2y);-i\omega_1r,-i(\omega_1+\omega_2))} \nonumber \\
	\times \frac{\gamma^{(2)}(-i(a_i+z+2\mu)-i\omega_2(r-(u_i+y));-i\omega_2r,-i(\omega_1+\omega_2))}{\gamma^{(2)}(\mp2i(z+\mu)-i\omega_2(r\mp(2y));-i\omega_2r,-i(\omega_1+\omega_2))}  \nonumber \\
	\prod_{i=4}^6\gamma^{(2)}(-i(a_i-z-2\mu)-i\omega_1(u_i- y);-i\omega_1r,-i(\omega_1+\omega_2)) \nonumber
	\\
	\times \gamma^{(2)}(-i(a_i-z-2\mu)-i\omega_2(r-(u_i- y));-i\omega_2r,-i(\omega_1+\omega_2))\Bigg] \nonumber \;.
	\end{align}
	Main idea is to transform the gauge group from $SU(2)$ to $U(1)$. In order to achieve this goal we will use the asymptotic relations of the special function $\gamma^{(2)}(z;\omega_1,\omega_2)$. Furthermore, as an outcome we will observe that there is a transformation in flavor group as well. After the process we also have the following right hand side
	\begin{align}
	= \prod_{i=1}^3\prod_{j=4}^6\gamma^{(2)}(-i(a_i+ a_j)-i\omega_1(u_i+ u_j);-i\omega_1r,-i(\omega_1+\omega_2))\nonumber \\ \gamma^{(2)}(-i(a_i+ a_j)-i\omega_2(r-(u_i+ u_j));-i\omega_2r,-i(\omega_1+\omega_2)) \nonumber \\   \Big[\prod_{1\leq i<j\leq 3}
	\gamma^{(2)}(-i(a_i+a_j+2\mu)-i\omega_1(u_i+u_j);-i\omega_1r,-i(\omega_1+\omega_2)) \nonumber \\\gamma^{(2)}(-i(a_i+a_j+2\mu)-i\omega_2(r-(u_i+u_j));-i\omega_2r,-i(\omega_1+\omega_2)) \nonumber \\
	\gamma^{(2)}(-i(a_{i+3}+a_{j+3}-2\mu)-i\omega_1(u_{i+3}+u_{j+3});-i\omega_1r,-i(\omega_1+\omega_2)) \nonumber \\\gamma^{(2)}(-i(a_{i+3}+a_{j+3}-2\mu)-i\omega_2(r-(u_{i+3}+u_{j+3}));-i\omega_2r,-i(\omega_1+\omega_2))\Big] \; .
	\end{align}
	Here, from the asymptotic relations which hyperbolic gamma function satisfy, each term in the brackets behaves in a particular way,

	\begin{align}
	e^{(\frac{2i\pi}{r}(\mu +z)(\omega_1^{-1}+\omega_2^{-1})+2i\pi y)}\gamma^{(2)}(\mp 2i(z+\mu)\mp i\omega_1(2y);-i\omega_1r,-i(\omega_1+\omega_2))^{-1}\nonumber\\\gamma^{(2)}(\mp 2i(z+\mu)\mp i\omega_2(r-2y);-i\omega_2r,-i(\omega_1+\omega_2))^{-1}\xrightarrow{\mu\to\infty} 1 \; ,
	\end{align}
and	
	\begin{align} \nonumber
	\prod_{i=1}^3~ & \gamma^{(2)}(-i(a_i+z+2\mu)-i\omega_1(u_i+y);-i\omega_1r,-i(\omega_1+\omega_2)) \nonumber \\
	\times~~~ & \gamma^{(2)}(-i(a_i+z+2\mu)-i\omega_2(r-(u_i+y));-i\omega_2r,-i(\omega_1+\omega_2))  \nonumber \\
	\times\prod_{i=4}^6~ & \gamma^{(2)}(-i(a_i-z-2\mu)-i\omega_1(u_i- y);-i\omega_1r,-i(\omega_1+\omega_2)) \nonumber
	\\
	\times~~~ & \gamma^{(2)}(-i(a_i-z-2\mu)-i\omega_2(r-(u_i- y));-i\omega_2r,-i(\omega_1+\omega_2))\nonumber  \\  
	\xrightarrow{\mu\to\infty}  \prod_{i=1}^3 ~ & e^{\left[B_{2,2}(-i(a_i+z+2\mu)-i\omega_1(u_i+y)) +B_{2,2}(-i(a_i+z+2\mu)-i\omega_2(r-(u_i+y)))\right]}\nonumber\\
	\times\prod_{i=4}^6 ~ & e^{\left[-B_{2,2}(-i(a_{i}-z-2\mu)-i\omega_1(u_{i}-y))-B_{2,2}(-i(a_{i}-z-2\mu)-i\omega_2(r-(u_{i}-y)))\right]}(1+o(1))  \; ,
	\end{align}
where we use shorthand notation as	$B_{2,2}(z)=B_{2,2}(z;-i\omega_{1,2} r,-i(\omega_1+\omega_2))$. At the right hand side,
	\begin{align} \nonumber
	\prod_{1\leq i<j\leq 3} &
	\gamma^{(2)}(-i(a_i+a_j+2\mu)-i\omega_1(u_i+u_j);-i\omega_1r,-i(\omega_1+\omega_2)) \nonumber \\\times ~~~& \gamma^{(2)}(-i(a_i+a_j+2\mu)-i\omega_2(r-(u_i+u_j));-i\omega_2r,-i(\omega_1+\omega_2)) \nonumber \\
	\times~~~ & \gamma^{(2)}(-i(a_{i+3}+a_{j+3}-2\mu)-i\omega_1(u_{i+3}+u_{j+3});-i\omega_1r,-i(\omega_1+\omega_2)) \nonumber \\\times ~~~ & \gamma^{(2)}(-i(a_{i+3}+a_{j+3}-2\mu)-i\omega_2(r-(u_{i+3}+u_{j+3}));-i\omega_2r,-i(\omega_1+\omega_2)) \nonumber\\
	\xrightarrow{\mu\to\infty}
	\prod_{1\leq i<j\leq3} & e^{\left[B_{2,2}(-i(a_i+a_j+2\mu)-i\omega_1(u_i+u_j))+B_{2,2}(-i(a_i+a_j+2\mu)-i\omega_2(r-(u_i+u_j)))\right]}\nonumber\\
	\times \prod_{4\leq i<j\leq6}  & e^{\left[-B_{2,2}(-i(a_{i}+a_{j}-2\mu)-i\omega_1(u_{i}+u_{j}))-B_{2,2}(-i(a_{i}+a_{j}-2\mu)-i\omega_2(r-(u_{i}+u_{j})))\right]}(1+o(1)) \; .
	\end{align}

	Hence, after the reduction of integration we rename $a_{i+3}=b_i$ and $u_{i+3}=v_i$, obtain (\ref{mainintegraltext})
	
	\begin{align}\label{newpentagon}
	\frac{1}{r\sqrt{-\omega_1\omega_2}}\sum_{y=0}^{[ r/2 ]}\epsilon (y) e^{\frac{\pi iC}{2}}\int _{-\infty}^{\infty} dz\prod_{i=1}^3 & \gamma^{(2)}(-i(a_i-z)-i\omega_1(u_i- y);-i\omega_1r,-i(\omega_1+\omega_2)) \nonumber \\
	\times & \gamma^{(2)}(-i(a_i-z)-i\omega_2(r-(u_i- y));-i\omega_2r,-i(\omega_1+\omega_2)) \nonumber \\
	\times & \gamma^{(2)}(-i(b_i+z)-i\omega_1(v_i+ y);-i\omega_1r,-i(\omega_1+\omega_2)) \nonumber
	\\
	\times & \gamma^{(2)}(-i(b_i+z)-i\omega_2(r-(v_i+y));-i\omega_2r,-i(\omega_1+\omega_2))\nonumber\\
	= \prod_{i,j=1}^3 & \gamma^{(2)}(-i(a_i+ b_j)-i\omega_1(u_i+ v_j);-i\omega_1r,-i(\omega_1+\omega_2))\nonumber\\\times & \gamma^{(2)}(-i(a_i+ b_j)-i\omega_2(r-(u_i+ v_j));-i\omega_2r,-i(\omega_1+\omega_2))\;.
	\end{align}
	with the balancing conditions $\sum_{i=1}^3 a_i+b_i=\omega_1+\omega_2$ and $\sum_{i=1}^3 u_i+v_i=0$. The constant term is $C=-2y+(u_1+u_2+u_3-v_1-v_2-v_3)$.
	\section{Star-Triangle Relation for \texorpdfstring{$r=2$}{r = 2}}
	We start by introducing two different gamma functions
	\begin{align}
	\Gamma(z;q,p)=\prod_{i,j=0}^\infty\frac{1-z^{-1}p^{i+1}q^{j+1}}{1-zp^iq^j} \; ,
	\end{align}
	\begin{align}
	\Gamma(z;q,p,t)=\prod_{i,j,k=0}^\infty(1-z^{-1}p^{i+1}q^{j+1}t^{k+1})(1-zp^iq^jt^k)
	\end{align}
	for $\mid q\mid,\mid p\mid, \mid t\mid<1$ and $z\in\mathbb C^*$.
	Additionally, we have the following identity,
	\begin{align}
	\Gamma(qz;p,q,t)=\Gamma(z;p,t)\Gamma(z;p,q,t) \;.
	\label{gamma_identity} 
	\end{align}
	We use the asymptotic relation between $\gamma^{(2)}(z;\omega_1,\omega_2)$ and $\Gamma(z;q,p)$ given as below
\begin{equation}
\Gamma(e^{2 \pi \textup{i} v z};e^{2 \pi \textup{i} v \omega_1}, e^{2 \pi
	\textup{i} v \omega_2}) \stackrel[{v \rightarrow 0}]{}{=} e^{-\pi
	\textup{i}(2z-(\omega_1+\omega_2))/24 v\omega_1\omega_2} \gamma^{(2)}(z;\omega_1,\omega_2) \; ,
\end{equation}
	For a particular asymptotic relation
	\begin{align}
	\Gamma(zq^y;q^r,qp)\Gamma(zp^{r-y},p^r,qp)\underset{v \to 0}{=}~~~~~~~~~~~~~~~~~~~~~~~~~~~~~~\nonumber \\ \gamma^{(2)}(-iz-i\omega_2(r-y);-i\omega_2r,-i\omega_1-i\omega_2)\gamma^{(2)}(-iz-i\omega_1y;-i\omega_1r,-i\omega_1-i\omega_2 )
	\end{align}
	where $q=e^{2\pi iv\omega_1}$ and $p=e^{2\pi iv\omega_2}$, we apply (\ref{gamma_identity})
	\begin{align}
	\Gamma(zq^y;q^r,qp)\Gamma(zp^{r-y},p^r,qp)= & \frac{ \Gamma(p^rzq^y;q^r,p^r,qp)}{ \Gamma(zq^y;q^r,p^r,qp)}\frac{\Gamma(q^rzp^{r-y};q^r ,p^r,qp)}{\Gamma(zp^{r-y};q^r,p^r,qp)} \\
	= & \frac{ \Gamma(zp^{r-y}(qp)^y;q^r,p^r,qp)}{ \Gamma(zp^{r-y};q^r,p^r,qp)}\frac{\Gamma(q^yz(qp)^{r-y};q^r,p^r,qp)}{\Gamma(zq^y;q^r,p^r,qp)} \\
	= & \Gamma(zp^{r-y}(qp)^{y-1};q^r,p^r)\Gamma(q^yz(qp)^{r-y-1};q^r,p^r) \; .
	\end{align}
	If we consider to use the identity in (\ref{gamma_identity}) only once, we observe the following asymptotic relation
	\begin{align} \nonumber
	\Gamma(zp^{r-y}(qp)^{y-1};q^r,p^r)\Gamma(q^yz(qp)^{r-y-1};q^r,p^r)\underset{v \to 0}{=} \\ \nonumber  \gamma^{(2)}(-iz-i\omega_1(r-1)-i\omega_2(r-y-1);-i\omega_1r,-i\omega_2r)\nonumber \\ \times \gamma^{(2)}(-iz-i\omega_2(r-1)-i\omega_1(y-1);-i\omega_1r,-i\omega_2r)  \; .
	\end{align}
	Moreover, for $y<r$ we use the identity several times and obtain the following form
	\begin{align}
	\Gamma(zq^y;q^r,qp)\Gamma(zp^{r-y},p^r,qp) & = \prod_{k=0}^{y-1}\Gamma(p^{r-y}z(qp)^k;q^r,p^r)\prod_{s=0}^{r-y-1}\Gamma(q^yz(qp)^s;q^r,p^r) \\
	& \underset{v \to 0}{=}\prod_{k=0}^{y-1}\gamma^{(2)}(-iz-i\omega_2(r-y+k)-ik\omega_1;-2i\omega_1,-2i\omega_2)\nonumber \\ & \times\prod_{s=0}^{r-y-1}\gamma^{(2)}(-iz-i\omega_1(y+s)-is\omega_2;-2i\omega_1,-2i\omega_2) \; .
	\label{limit_relation}
	\end{align}
	For r=2, we calculate the cases y=0 and y=1 from (\ref{limit_relation}) and derive the relation between $\gamma^{(2)}(z;\omega_1,\omega_2)$ pairs as follows
	\begin{align}
	\gamma^{(2)}(-iz-i\omega_2(2-y);-2i\omega_2,-i\omega_1-i\omega_2)\gamma^{(2)}(-iz-i\omega_1y;-2i\omega_1,-i\omega_1-i\omega_2 ) \nonumber \\
	=\gamma^{(2)}(-iz-i\omega_2-i(1-y)\omega_1;-2i\omega_1,-2i\omega_2)\gamma^{(2)}(-iz-iy\omega_1;-2i\omega_1,-2i\omega_2)
	\label{schomerous} \;.
	\end{align}
	Furthermore, we use (\ref{schomerous}) to rewrite the (\ref{mainintegral}) explicitly,
	\begin{align} \nonumber
	\frac{1}{2\sqrt{-\omega_1\omega_2}}\sum_{y=0}^{1}e^{\frac{\pi i}{2}(2y-\sum_i( u_i-v_i))} \int _{-\infty}^{\infty} dz
	\prod_{i=1}^3 & \gamma^{(2)}(-i(a_i-z)-i\omega_1(u_i- y);-2i\omega_1,-2i\omega_2) \nonumber \\
	\times & \gamma^{(2)}(-i(a_i-z)-i\omega_2-i\omega_2(1-(u_i- y));-2i\omega_2,-2i\omega_1) \nonumber \\ \times
	& \gamma^{(2)}(-i(b_i+z)-i\omega_1(v_i+ y);-2i\omega_1,-2i\omega_2) \nonumber
	\\
	\times & \gamma^{(2)}(-i(b_i+z)-i\omega_2-i\omega_1(1-(v_i+y));-2i\omega_2,-2i\omega_1) \nonumber \\
	=\prod_{i,j=1}^3 & \gamma^{(2)}(-i(a_i+ b_j)-i\omega_1(u_i+ v_j);-2i\omega_1,-2i\omega_2)\nonumber \\\times & \gamma^{(2)}(-i(a_i+ b_j)-i\omega_2-i\omega_1(1-(u_i+ v_j));-2i\omega_2,-2i\omega_1) \; .
	\end{align}
	Than we identify $-2i\omega_1$ with b, $-2i\omega_2$ with 1/b, redefine $u_i=\mu_i$, $v_i=1-\nu_i$ and all coefficients without -2i multiplier. 
	Thus, the integral identity takes the form
	\begin{align}
	\sum_{\nu=0}^{1}(-1)^{\frac{2\nu-3-\sum_i( \mu_i-\nu_i)}{2}} \int _{-\infty}^{\infty} \frac{dx}{2i}\prod_{i=1}^3 & \gamma^{(2)}(\frac{a_i+x+b(1-\nu_i-\nu)}{2};b,b^{-1}) \nonumber \\
	\times~ & \gamma^{(2)}(\frac{a_i+x+b^{-1}+b(\nu_i+ \nu)}{2};b,b^{-1}) \nonumber \\
	\times~ & \gamma^{(2)}(\frac{b_i-x+b(-\mu_i+ \nu)}{2};b,b^{-1}) \nonumber
	\\
	\times ~ &\gamma^{(2)}(\frac{(b_i-x+b^{-1}+b(1+\mu_i-\nu))}{2};b,b^{-1}) \nonumber \\
	=\prod_{i,j=1}^3 \gamma^{(2)}(\frac{a_i+ b_j+b(1-\nu_i-\mu_j)}{2};b,b^{-1}) ~ & \gamma^{(2)}(\frac{a_i+ b_j+b^{-1}+b(\nu_i+ \mu_j))}{2};b,b^{-1}) \; .
	\end{align}
	This is exactly the integral identity (\ref{r=2iden}). 


\begin{thebibliography}{10}
	
	\bibitem{Gahramanov:2017ysd}
	I.~Gahramanov and S.~Jafarzade, ``{Integrable lattice spin models from
		supersymmetric dualities},''
	\href{http://dx.doi.org/10.1134/S1547477118060079}{{\em Phys. Part. Nucl.
			Lett.} {\bfseries 15} no.~6, (2018) 650--667},
	\href{http://arxiv.org/abs/1712.09651}{{\ttfamily arXiv:1712.09651
			[math-ph]}}.
	
	\bibitem{Yamazaki:2018xbx}
	M.~Yamazaki, ``{Integrability As Duality: The Gauge/YBE Correspondence},''
	\href{http://dx.doi.org/10.1016/j.physrep.2020.01.006}{{\em Phys. Rept.}
		{\bfseries 859} (2020) 1--20},
	\href{http://arxiv.org/abs/1808.04374}{{\ttfamily arXiv:1808.04374
			[hep-th]}}.
	
	\bibitem{Spiridonov:2010em}
	V.~Spiridonov, ``{Elliptic beta integrals and solvable models of statistical
		mechanics},'' {\em Contemp. Math.} {\bfseries 563} (2012) 181--211,
	\href{http://arxiv.org/abs/1011.3798}{{\ttfamily arXiv:1011.3798 [hep-th]}}.
	
	\bibitem{Kels:2015bda}
	A.~P. Kels, ``{New solutions of the star--triangle relation with discrete and
		continuous spin variables},''
	\href{http://dx.doi.org/10.1088/1751-8113/48/43/435201}{{\em J. Phys. A}
		{\bfseries 48} no.~43, (2015) 435201},
	\href{http://arxiv.org/abs/1504.07074}{{\ttfamily arXiv:1504.07074
			[math-ph]}}.
	
	\bibitem{Gahramanov:2015cva}
	I.~Gahramanov and V.~Spiridonov, ``{The star-triangle relation and 3d
		superconformal indices},''
	\href{http://dx.doi.org/10.1007/JHEP08(2015)040}{{\em JHEP} {\bfseries 08}
		(2015) 040}, \href{http://arxiv.org/abs/1505.00765}{{\ttfamily
			arXiv:1505.00765 [hep-th]}}.
	
	\bibitem{Kels:2017toi}
	A.~P. Kels and M.~Yamazaki, ``{Elliptic hypergeometric sum/integral
		transformations and supersymmetric lens index},''
	\href{http://dx.doi.org/10.3842/SIGMA.2018.013}{{\em SIGMA} {\bfseries 14}
		(2018) 013}, \href{http://arxiv.org/abs/1704.03159}{{\ttfamily
			arXiv:1704.03159 [math-ph]}}.
	
	\bibitem{Jafarzade:2017fsc}
	S.~Jafarzade and Z.~Nazari, ``{A New Integrable Ising-type Model from 2d
		$\mathcal{N}$=(2,2) Dualities},''
	\href{http://arxiv.org/abs/1709.00070}{{\ttfamily arXiv:1709.00070
			[hep-th]}}.
	
	\bibitem{de-la-Cruz-Moreno:2020xop}
	J.~de-la Cruz-Moreno and H.~García-Compeán, ``{Star-triangle type relations
		from $2d$ $\mathcal{N}=(0,2)$ $USp(2N)$ dualities},''
	\href{http://arxiv.org/abs/2008.02419}{{\ttfamily arXiv:2008.02419
			[hep-th]}}.
	
	\bibitem{Gahramanov:2016ilb}
	I.~Gahramanov and A.~P. Kels, ``{The star-triangle relation, lens partition
		function, and hypergeometric sum/integrals},''
	\href{http://dx.doi.org/10.1007/JHEP02(2017)040}{{\em JHEP} {\bfseries 02}
		(2017) 040}, \href{http://arxiv.org/abs/1610.09229}{{\ttfamily
			arXiv:1610.09229 [math-ph]}}.
	
	\bibitem{Baxter:1982zz}
	R.~Baxter, {\em {Exactly solved models in statistical mechanics}}.
	\newblock 1982.
	
	\bibitem{Baxter:1997tn}
	R.~Baxter, ``{Star-triangle and star-star relations in statistical
		mechanics},'' \href{http://dx.doi.org/10.1142/S0217979297000058}{{\em Int. J.
			Mod. Phys. B} {\bfseries 11} (1997) 27--37}.
	
	\bibitem{Gang:2019juz}
	D.~Gang, ``{Chern-Simons Theory on $L(p,q)$ Lens Spaces and Localization},''
	\href{http://dx.doi.org/10.3938/jkps.74.1119}{{\em J. Korean Phys. Soc.}
		{\bfseries 74} no.~12, (2019) 1119--1128},
	\href{http://arxiv.org/abs/0912.4664}{{\ttfamily arXiv:0912.4664 [hep-th]}}.
	
	\bibitem{Benini:2011nc}
	F.~Benini, T.~Nishioka, and M.~Yamazaki, ``{4d Index to 3d Index and 2d
		TQFT},'' \href{http://dx.doi.org/10.1103/PhysRevD.86.065015}{{\em Phys. Rev.
			D} {\bfseries 86} (2012) 065015},
	\href{http://arxiv.org/abs/1109.0283}{{\ttfamily arXiv:1109.0283 [hep-th]}}.
	
	\bibitem{Imamura:2012rq}
	Y.~Imamura and D.~Yokoyama, ``{$S^3/Z_n$ partition function and dualities},''
	\href{http://dx.doi.org/10.1007/JHEP11(2012)122}{{\em JHEP} {\bfseries 11}
		(2012) 122}, \href{http://arxiv.org/abs/1208.1404}{{\ttfamily arXiv:1208.1404
			[hep-th]}}.
	
	\bibitem{Imamura:2013qxa}
	Y.~Imamura, H.~Matsuno, and D.~Yokoyama, ``{Factorization of the
		$S^3/\mathbb{Z}_n$ partition function},''
	\href{http://dx.doi.org/10.1103/PhysRevD.89.085003}{{\em Phys. Rev. D}
		{\bfseries 89} no.~8, (2014) 085003},
	\href{http://arxiv.org/abs/1311.2371}{{\ttfamily arXiv:1311.2371 [hep-th]}}.
	
	\bibitem{Nieri:2015yia}
	F.~Nieri and S.~Pasquetti, ``{Factorisation and holomorphic blocks in 4d},''
	\href{http://dx.doi.org/10.1007/JHEP11(2015)155}{{\em JHEP} {\bfseries 11}
		(2015) 155}, \href{http://arxiv.org/abs/1507.00261}{{\ttfamily
			arXiv:1507.00261 [hep-th]}}.
	
	\bibitem{Alday:2012au}
	L.~F. Alday, M.~Fluder, and J.~Sparks, ``{The Large N limit of M2-branes on
		Lens spaces},'' \href{http://dx.doi.org/10.1007/JHEP10(2012)057}{{\em JHEP}
		{\bfseries 10} (2012) 057}, \href{http://arxiv.org/abs/1204.1280}{{\ttfamily
			arXiv:1204.1280 [hep-th]}}.
	
	\bibitem{Yamazaki:2013fva}
	M.~Yamazaki, ``{Four-dimensional superconformal index reloaded},''
	\href{http://dx.doi.org/10.1007/s11232-013-0012-6}{{\em Theor. Math. Phys.}
		{\bfseries 174} (2013) 154--166}.
	
	\bibitem{Eren:2019ibl}
	E.~Eren, I.~Gahramanov, S.~Jafarzade, and G.~Mogol, ``{Gamma function solutions
		to the star-triangle equation},''
	\href{http://arxiv.org/abs/1912.12271}{{\ttfamily arXiv:1912.12271
			[math-ph]}}.
	
	\bibitem{Honda:2016vmv}
	M.~Honda, ``{How to resum perturbative series in 3d N=2 Chern-Simons matter
		theories},'' \href{http://dx.doi.org/10.1103/PhysRevD.94.025039}{{\em Phys.
			Rev. D} {\bfseries 94} no.~2, (2016) 025039},
	\href{http://arxiv.org/abs/1604.08653}{{\ttfamily arXiv:1604.08653
			[hep-th]}}.
	
	\bibitem{Nedelin:2016gwu}
	A.~Nedelin, F.~Nieri, and M.~Zabzine, ``{$q$-Virasoro modular double and 3d
		partition functions},''
	\href{http://dx.doi.org/10.1007/s00220-017-2882-1}{{\em Commun. Math. Phys.}
		{\bfseries 353} no.~3, (2017) 1059--1102},
	\href{http://arxiv.org/abs/1605.07029}{{\ttfamily arXiv:1605.07029
			[hep-th]}}.
	
	\bibitem{Bazhanov:2007mh}
	V.~V. Bazhanov, V.~V. Mangazeev, and S.~M. Sergeev, ``{Faddeev-Volkov solution
		of the Yang-Baxter equation and discrete conformal symmetry},''
	\href{http://dx.doi.org/10.1016/j.nuclphysb.2007.05.013}{{\em Nucl. Phys. B}
		{\bfseries 784} (2007) 234--258},
	\href{http://arxiv.org/abs/hep-th/0703041}{{\ttfamily arXiv:hep-th/0703041}}.
	
	\bibitem{Bazhanov:2007vg}
	V.~V. Bazhanov, V.~V. Mangazeev, and S.~M. Sergeev, ``{Exact solution of the
		Faddeev-Volkov model},''
	\href{http://dx.doi.org/10.1016/j.physleta.2007.10.053}{{\em Phys. Lett. A}
		{\bfseries 372} (2008) 1547--1550},
	\href{http://arxiv.org/abs/0706.3077}{{\ttfamily arXiv:0706.3077
			[cond-mat.stat-mech]}}.
	
	\bibitem{Jafarzade:2018yei}
	S.~Jafarzade, ``{New Pentagon Identities Revisited},''
	\href{http://dx.doi.org/10.1088/1742-6596/1194/1/012054}{{\em J. Phys. Conf.
			Ser.} {\bfseries 1194} no.~1, (2019) 012054},
	\href{http://arxiv.org/abs/1812.01325}{{\ttfamily arXiv:1812.01325
			[math-ph]}}.
	
	\bibitem{Hadasz:2013bwa}
	L.~Hadasz, M.~Pawelkiewicz, and V.~Schomerus, ``{Self-dual Continuous Series of
		Representations for U\_q(sl(2)) and U\_q(osp(1|2))},''
	\href{http://dx.doi.org/10.1007/JHEP10(2014)091}{{\em JHEP} {\bfseries 10}
		(2014) 091}, \href{http://arxiv.org/abs/1305.4596}{{\ttfamily arXiv:1305.4596
			[hep-th]}}.
	
	\bibitem{Sarkissian:2018ppc}
	G.~Sarkissian and V.~P. Spiridonov, ``{From rarefied elliptic beta integral to
		parafermionic star-triangle relation},''
	\href{http://dx.doi.org/10.1007/JHEP10(2018)097}{{\em JHEP} {\bfseries 10}
		(2018) 097}, \href{http://arxiv.org/abs/1809.00493}{{\ttfamily
			arXiv:1809.00493 [hep-th]}}.
	
	\bibitem{Alday:2013lba}
	L.~F. Alday, D.~Martelli, P.~Richmond, and J.~Sparks, ``{Localization on
		Three-Manifolds},'' \href{http://dx.doi.org/10.1007/JHEP10(2013)095}{{\em
			JHEP} {\bfseries 10} (2013) 095},
	\href{http://arxiv.org/abs/1307.6848}{{\ttfamily arXiv:1307.6848 [hep-th]}}.
	
	\bibitem{Willett:2016adv}
	B.~Willett, ``{Localization on three-dimensional manifolds},''
	\href{http://dx.doi.org/10.1088/1751-8121/aa612f}{{\em J. Phys. A} {\bfseries
			50} no.~44, (2017) 443006}, \href{http://arxiv.org/abs/1608.02958}{{\ttfamily
			arXiv:1608.02958 [hep-th]}}.
	
	\bibitem{Benini:2013yva}
	F.~Benini and W.~Peelaers, ``{Higgs branch localization in three dimensions},''
	\href{http://dx.doi.org/10.1007/JHEP05(2014)030}{{\em JHEP} {\bfseries 05}
		(2014) 030}, \href{http://arxiv.org/abs/1312.6078}{{\ttfamily arXiv:1312.6078
			[hep-th]}}.
	
	\bibitem{van2007hyperbolic}
	F.~van~de Bult {\em et~al.}, ``Hyperbolic hypergeometric functions,'' {\em
		Ph.D. Thesis, University of Amsterdam, Amsterdam Netherlands} (2007) .
	
	\bibitem{Amariti:2015vwa}
	A.~Amariti, ``{Integral identities for 3d dualities with SP(2N) gauge
		groups},'' \href{http://arxiv.org/abs/1509.02199}{{\ttfamily arXiv:1509.02199
			[hep-th]}}.
	
	\bibitem{Spiridonov:2016uae}
	V.~Spiridonov, ``{Rarefied elliptic hypergeometric functions},''
	\href{http://dx.doi.org/10.1016/j.aim.2018.04.014}{{\em Adv. Math.}
		{\bfseries 331} (2018) 830--873},
	\href{http://arxiv.org/abs/1609.00715}{{\ttfamily arXiv:1609.00715
			[math.CA]}}.
	
	\bibitem{Cordova:2016jlu}
	C.~Cordova, B.~Heidenreich, A.~Popolitov, and S.~Shakirov, ``{Orbifolds and
		Exact Solutions of Strongly-Coupled Matrix Models},''
	\href{http://dx.doi.org/10.1007/s00220-017-3072-x}{{\em Commun. Math. Phys.}
		{\bfseries 361} no.~3, (2018) 1235--1274},
	\href{http://arxiv.org/abs/1611.03142}{{\ttfamily arXiv:1611.03142
			[hep-th]}}.
	
	\bibitem{Spiridonov:2008zr}
	V.~Spiridonov and G.~Vartanov, ``{Superconformal indices for N = 1 theories
		with multiple duals},''
	\href{http://dx.doi.org/10.1016/j.nuclphysb.2009.08.022}{{\em Nucl. Phys. B}
		{\bfseries 824} (2010) 192--216},
	\href{http://arxiv.org/abs/0811.1909}{{\ttfamily arXiv:0811.1909 [hep-th]}}.
	
	\bibitem{Gahramanov:2013xsa}
	I.~Gahramanov and G.~Vartanov, ``{Extended global symmetries for 4D $N$ = 1
		SQCD theories},''
	\href{http://dx.doi.org/10.1088/1751-8113/46/28/285403}{{\em J. Phys. A}
		{\bfseries 46} (2013) 285403},
	\href{http://arxiv.org/abs/1303.1443}{{\ttfamily arXiv:1303.1443 [hep-th]}}.
	
	\bibitem{Aharony:2013dha}
	O.~Aharony, S.~S. Razamat, N.~Seiberg, and B.~Willett, ``{3d dualities from 4d
		dualities},'' \href{http://dx.doi.org/10.1007/JHEP07(2013)149}{{\em JHEP}
		{\bfseries 07} (2013) 149}, \href{http://arxiv.org/abs/1305.3924}{{\ttfamily
			arXiv:1305.3924 [hep-th]}}.
	
	\bibitem{Karch:1997ux}
	A.~Karch, ``{Seiberg duality in three-dimensions},''
	\href{http://dx.doi.org/10.1016/S0370-2693(97)00598-4}{{\em Phys. Lett. B}
		{\bfseries 405} (1997) 79--84},
	\href{http://arxiv.org/abs/hep-th/9703172}{{\ttfamily arXiv:hep-th/9703172}}.
	
	\bibitem{Dolan:2008qi}
	F.~Dolan and H.~Osborn, ``{Applications of the Superconformal Index for
		Protected Operators and q-Hypergeometric Identities to N=1 Dual Theories},''
	\href{http://dx.doi.org/10.1016/j.nuclphysb.2009.01.028}{{\em Nucl. Phys. B}
		{\bfseries 818} (2009) 137--178},
	\href{http://arxiv.org/abs/0801.4947}{{\ttfamily arXiv:0801.4947 [hep-th]}}.
	
	\bibitem{Hama:2011ea}
	N.~Hama, K.~Hosomichi, and S.~Lee, ``{SUSY Gauge Theories on Squashed
		Three-Spheres},'' \href{http://dx.doi.org/10.1007/JHEP05(2011)014}{{\em JHEP}
		{\bfseries 05} (2011) 014}, \href{http://arxiv.org/abs/1102.4716}{{\ttfamily
			arXiv:1102.4716 [hep-th]}}.
	
	\bibitem{Kashaev:2012cz}
	R.~Kashaev, F.~Luo, and G.~Vartanov, ``{A TQFT of Turaev--Viro Type on Shaped
		Triangulations},'' \href{http://dx.doi.org/10.1007/s00023-015-0427-8}{{\em
			Annales Henri Poincare} {\bfseries 17} no.~5, (2016) 1109--1143},
	\href{http://arxiv.org/abs/1210.8393}{{\ttfamily arXiv:1210.8393 [math.QA]}}.
	
	\bibitem{Gahramanov:2013rda}
	I.~Gahramanov and H.~Rosengren, ``{A new pentagon identity for the tetrahedron
		index},'' \href{http://dx.doi.org/10.1007/JHEP11(2013)128}{{\em JHEP}
		{\bfseries 11} (2013) 128}, \href{http://arxiv.org/abs/1309.2195}{{\ttfamily
			arXiv:1309.2195 [hep-th]}}.
	
	\bibitem{Gahramanov:2014ona}
	I.~Gahramanov and H.~Rosengren, ``{Integral pentagon relations for 3d
		superconformal indices},''
	\href{http://dx.doi.org/10.1090/pspum/093/01569}{{\em Proc. Symp. Pure Math.}
		{\bfseries 93} (2016) 165--173},
	\href{http://arxiv.org/abs/1412.2926}{{\ttfamily arXiv:1412.2926 [hep-th]}}.
	
	\bibitem{Gahramanov:2016wxi}
	I.~Gahramanov and H.~Rosengren, ``{Basic hypergeometry of supersymmetric
		dualities},'' \href{http://dx.doi.org/10.1016/j.nuclphysb.2016.10.004}{{\em
			Nucl. Phys. B} {\bfseries 913} (2016) 747--768},
	\href{http://arxiv.org/abs/1606.08185}{{\ttfamily arXiv:1606.08185
			[hep-th]}}.
	
	\bibitem{pachner1991pl}
	U.~Pachner, ``P.l. homeomorphic manifolds are equivalent by elementary
	shellings,'' \href{http://dx.doi.org/10.1016/S0195-6698(13)80080-7}{{\em
			European journal of Combinatorics} {\bfseries 12} no.~2, (1991) 129--145}.
	
	\bibitem{Dimofte:2011ju}
	T.~Dimofte, D.~Gaiotto, and S.~Gukov, ``{Gauge Theories Labelled by
		Three-Manifolds},'' \href{http://dx.doi.org/10.1007/s00220-013-1863-2}{{\em
			Commun. Math. Phys.} {\bfseries 325} (2014) 367--419},
	\href{http://arxiv.org/abs/1108.4389}{{\ttfamily arXiv:1108.4389 [hep-th]}}.
	
	\bibitem{Benvenuti:2016wet}
	S.~Benvenuti and S.~Pasquetti, ``{3d $ \mathcal{N} $ = 2 mirror symmetry,
		pq-webs and monopole superpotentials},''
	\href{http://dx.doi.org/10.1007/JHEP08(2016)136}{{\em JHEP} {\bfseries 08}
		(2016) 136}, \href{http://arxiv.org/abs/1605.02675}{{\ttfamily
			arXiv:1605.02675 [hep-th]}}.
	
	\bibitem{Bozkurt:2018xno}
	D.~N. Bozkurt and I.~Gahramanov, ``{Pentagon identities arising in
		supersymmetric gauge theory computations},''
	\href{http://dx.doi.org/10.1134/S0040577919020028}{{\em Teor. Mat. Fiz.}
		{\bfseries 198} no.~2, (2019) 215--224},
	\href{http://arxiv.org/abs/1803.00855}{{\ttfamily arXiv:1803.00855
			[math-ph]}}.
	
	\bibitem{Alexandrov:2015xir}
	S.~Alexandrov and B.~Pioline, ``{Theta series, wall-crossing and quantum
		dilogarithm identities},''
	\href{http://dx.doi.org/10.1007/s11005-016-0857-3}{{\em Lett. Math. Phys.}
		{\bfseries 106} no.~8, (2016) 1037--1066},
	\href{http://arxiv.org/abs/1511.02892}{{\ttfamily arXiv:1511.02892
			[hep-th]}}.
	
	\bibitem{kashaev2014euler}
	R.~Kashaev, ``Euler’s beta function and pentagon relations,''
	\href{http://dx.doi.org/https://doi.org/10.1007/s40306-014-0080-1}{{\em Acta
			Mathematica Vietnamica} {\bfseries 39} no.~4, (2014) 561--566}.
	
	\bibitem{Benini:2012ui}
	F.~Benini and S.~Cremonesi, ``{Partition Functions of ${\mathcal{N}=(2,2)}$
		Gauge Theories on S$^{2}$ and Vortices},''
	\href{http://dx.doi.org/10.1007/s00220-014-2112-z}{{\em Commun. Math. Phys.}
		{\bfseries 334} no.~3, (2015) 1483--1527},
	\href{http://arxiv.org/abs/1206.2356}{{\ttfamily arXiv:1206.2356 [hep-th]}}.
	
	\bibitem{Zamolodchikov:1980mb}
	A.~Zamolodchikov, ``{'Fishnet' Diagrams as a Completely Integrable System},''
	\href{http://dx.doi.org/10.1016/0370-2693(80)90547-X}{{\em Phys. Lett. B}
		{\bfseries 97} (1980) 63--66}.
	
	\bibitem{Bazhanov:2016ajm}
	V.~V. Bazhanov, A.~P. Kels, and S.~M. Sergeev, ``{Quasi-classical expansion of
		the star-triangle relation and integrable systems on quad-graphs},''
	\href{http://dx.doi.org/10.1088/1751-8113/49/46/464001}{{\em J. Phys. A}
		{\bfseries 49} no.~46, (2016) 464001},
	\href{http://arxiv.org/abs/1602.07076}{{\ttfamily arXiv:1602.07076
			[math-ph]}}.
	
	\bibitem{Kulish:1988gr}
	P.~Kulish, ``{Quantum Superalgebra Osp(2|1)},''
	\href{http://dx.doi.org/10.1007/BF01101123}{{\em J. Sov. Math.} {\bfseries
			54} (1989) 923--930}.
	
	\bibitem{Kulish:1989sv}
	P.~Kulish and N.~Reshetikhin, ``{Universal R matrix of the quantum superalgebra
		osp(2 | 1)},'' \href{http://dx.doi.org/10.1007/BF00401868}{{\em Lett. Math.
			Phys.} {\bfseries 18} (1989) 143--149}.
	
	\bibitem{Saleur:1989gj}
	H.~Saleur, ``{Quantum Osp(1,2) and Solutions of the Graded {Yang-Baxter}
		Equation},'' \href{http://dx.doi.org/10.1016/0550-3213(90)90433-E}{{\em Nucl.
			Phys. B} {\bfseries 336} (1990) 363--376}.
	
	\bibitem{Pawelkiewicz:2013wga}
	M.~Pawelkiewicz, V.~Schomerus, and P.~Suchanek, ``{The universal Racah-Wigner
		symbol for U$_q$(osp(1|2))},''
	\href{http://dx.doi.org/10.1007/JHEP04(2014)079}{{\em JHEP} {\bfseries 04}
		(2014) 079}, \href{http://arxiv.org/abs/1307.6866}{{\ttfamily arXiv:1307.6866
			[hep-th]}}.
	
	\bibitem{Poghosyan:2016kvd}
	H.~Poghosyan and G.~Sarkissian, ``{Comments on fusion matrix in N=1 super
		Liouville field theory},''
	\href{http://dx.doi.org/10.1016/j.nuclphysb.2016.05.023}{{\em Nucl. Phys. B}
		{\bfseries 909} (2016) 458--479},
	\href{http://arxiv.org/abs/1602.07476}{{\ttfamily arXiv:1602.07476
			[hep-th]}}.
	
	\bibitem{Teschner:2012em}
	J.~Teschner and G.~Vartanov, ``{6j symbols for the modular double, quantum
		hyperbolic geometry, and supersymmetric gauge theories},''
	\href{http://dx.doi.org/10.1007/s11005-014-0684-3}{{\em Lett. Math. Phys.}
		{\bfseries 104} (2014) 527--551},
	\href{http://arxiv.org/abs/1202.4698}{{\ttfamily arXiv:1202.4698 [hep-th]}}.
	
	\bibitem{Hadasz:2007wi}
	L.~Hadasz, ``{On the fusion matrix of the N=1 Neveu-Schwarz blocks},''
	\href{http://dx.doi.org/10.1088/1126-6708/2007/12/071}{{\em JHEP} {\bfseries
			12} (2007) 071}, \href{http://arxiv.org/abs/0707.3384}{{\ttfamily
			arXiv:0707.3384 [hep-th]}}.
	
	\bibitem{kashaev1997heisenberg}
	R.~M. Kashaev, ``Heisenberg double and the pentagon relation,'' in {\em St.
		Petersburg Math. J}, Citeseer.
	\newblock 1997.
	
	\bibitem{Faddeev:1999fe}
	L.~Faddeev, ``{Modular double of quantum group},'' in {\em {Conference Moshe
			Flato}}, pp.~149--156.
	\newblock 2000.
	\newblock \href{http://arxiv.org/abs/math/9912078}{{\ttfamily
			arXiv:math/9912078}}.
	
	\bibitem{hikami2001hyperbolic}
	K.~Hikami, ``Hyperbolic structure arising from a knot invariant,''
	\href{http://dx.doi.org/10.1142/s0217751x0100444x}{{\em International Journal
			of Modern Physics A} {\bfseries 16} no.~19, (2001) 3309--3333}.
	
	\bibitem{Andersen:2014aoa}
	J.~E. Andersen and R.~Kashaev, ``{Complex Quantum Chern-Simons},''
	\href{http://arxiv.org/abs/1409.1208}{{\ttfamily arXiv:1409.1208 [math.QA]}}.
	
	\bibitem{Hikami_2007}
	K.~Hikami, ``Generalized volume conjecture and the a-polynomials: The
	neumann–zagier potential function as a classical limit of the partition
	function,'' \href{http://dx.doi.org/10.1016/j.geomphys.2007.03.008}{{\em
			Journal of Geometry and Physics} {\bfseries 57} no.~9, (Aug, 2007)
		1895–1940}.
	
	\bibitem{hikami2014braiding}
	K.~Hikami and R.~Inoue, ``Braiding operator via quantum cluster algebra,''
	\href{http://dx.doi.org/10.1088/1751-8113/47/47/474006}{{\em J. Phys A}
		{\bfseries 47} no.~47, (2014) 474006}.
	
	\bibitem{Chan:2017qnw}
	C.-T. Chan, A.~Mironov, A.~Morozov, and A.~Sleptsov, ``{Orthogonal Polynomials
		in Mathematical Physics},''
	\href{http://dx.doi.org/10.1142/9789813233867\_0011}{{\em Rev.Math.Phys}
		{\bfseries 30} (2018) 1840005},
	\href{http://arxiv.org/abs/1712.03155}{{\ttfamily arXiv:1712.03155
			[hep-th]}}.
	
	\bibitem{Faddeev:1995nb}
	L.~Faddeev, ``{Discrete Heisenberg-Weyl group and modular group},''
	\href{http://dx.doi.org/10.1007/BF01872779}{{\em Lett. Math. Phys.}
		{\bfseries 34} (1995) 249--254},
	\href{http://arxiv.org/abs/hep-th/9504111}{{\ttfamily arXiv:hep-th/9504111}}.
	
	\bibitem{woronowicz2000quantum}
	S.~Woronowicz, ``Quantum exponential function,''
	\href{http://dx.doi.org/10.1142/S0129055X00000344}{{\em Reviews in
			Mathematical Physics} {\bfseries 12} no.~06, (2000) 873--920}.
	
	\bibitem{Kashaev:2015nya}
	R.~Kashaev, ``{The Yang--Baxter relation and gauge invariance},''
	\href{http://dx.doi.org/10.1088/1751-8113/49/16/164001}{{\em J. Phys. A}
		{\bfseries 49} no.~16, (2016) 164001},
	\href{http://arxiv.org/abs/1510.03043}{{\ttfamily arXiv:1510.03043
			[math-ph]}}.
	
\end{thebibliography}
\end{document}